\documentclass[12pt,preprint]{aastex}

\slugcomment{}

\shorttitle{CSTAR exoplanet candidates}
\shortauthors{Wang et al.}

\begin{document}



\title{Planetary Transit Candidates in the CSTAR Field: Analysis of the 2008 Data}


\author{Songhu Wang\altaffilmark{1}, Hui Zhang\altaffilmark{1}, Ji-Lin Zhou\altaffilmark{1}, Xu Zhou\altaffilmark{2,9}, Ming Yang\altaffilmark{1}, Lifan Wang\altaffilmark{3}, D. Bayliss\altaffilmark{4}, G. Zhou\altaffilmark{4}, M. C. B. Ashley\altaffilmark{5}, Zhou Fan\altaffilmark{2,9}, Long-Long Feng\altaffilmark{3}, Xuefei Gong\altaffilmark{6}, J. S. Lawrence\altaffilmark{5,7}, Huigen Liu\altaffilmark{1}, Qiang Liu\altaffilmark{2}, D. M. Luong-Van\altaffilmark{5}, Jun Ma\altaffilmark{2,9}, Zeyang Meng\altaffilmark{1}, J. W. V. Storey\altaffilmark{5}, R. A. Wittenmyer\altaffilmark{5}, Zhenyu Wu\altaffilmark{2,9}, Jun Yan\altaffilmark{2}, Huigen Yang\altaffilmark{8}, Ji Yang\altaffilmark{3}, Jiayi Yang\altaffilmark{1}, Xiangyan Yuan\altaffilmark{6}, Tianmeng Zhang\altaffilmark{2,9}, Zhenxi Zhu\altaffilmark{3}, Hu Zou\altaffilmark{2,9}}
\email{zhoujl@nju.edu.cn, zhouxu@bao.ac.cn}

\altaffiltext{1}{Department of Astronomy \& Key Laboratory of Modern Astronomy and Astrophysics in Ministry of Education, Nanjing University, Nanjing 210093, China; zhoujl@nju.edu.cn}
\altaffiltext{2}{National Astronomical Observatories, Chinese Academy of Sciences, Beijing 100012, China; zhouxu@bao.ac.cn}
\altaffiltext{3}{Purple Mountain Observatory, Chinese Academy of Sciences, Nanjing 210008, China}
\altaffiltext{4}{Research School of Astronomy and Astrophysics, Australian National University, Canberra, ACT 2611, Australia}
\altaffiltext{5}{School of Physics, University of New South Wales, NSW 2052, Australia}
\altaffiltext{6}{Nanjing Institute of Astronomical Optics and Technology, Nanjing 210042, China }
\altaffiltext{7}{Australian Astronomical Observatory, NSW 1710, Australia }
\altaffiltext{8}{Polar Research Institute of China, Pudong, Shanghai 200136, China}
\altaffiltext{9}{Key Laboratory of Optical Astronomy, National Astronomical Observatories, Chinese Academy of Sciences, Beijing 100012, China}

\begin{abstract}
The Chinese Small Telescope ARray (CSTAR) is a group of four identical, fully automated, static $14.5\,\rm{cm}$ telescopes.
CSTAR is located at Dome A, Antarctica and covers $20\,\rm deg^2$ of sky around the South Celestial Pole.
The installation is designed to provide high-cadence photometry for the purpose of monitoring the quality of the astronomical observing conditions at Dome A and detecting transiting exoplanets. CSTAR has been operational since 2008, and has taken a rich and high-precision photometric data set of 10,690 stars. In the first observing season, we obtained 291,911 qualified science frames with 20-second integrations in the $i$-band. Photometric precision reaches $\sim 4\,\rm{mmag}$ at 20-second cadence at $i=7.5$, and is $\sim 20\,\rm{mmag}$ at $i=12$. Using robust detection methods, ten promising exoplanet candidates were found. Four of these were found to be giants using spectroscopic follow-up. All of these transit candidates are presented here along with the discussion of their detailed properties as well as the follow-up observations.
\end{abstract}




\section{Introduction}

The detection and study of exoplanets is one of the most exciting and fastest growing fields in astrophysics. At the present time, several different detection methods have yielded success. Two of most productive methods among them have been the radial velocity method and the transit method. Even though among the confirmed exoplanets, the radial velocity method has been more productive, the transit method also has its own advantages.
The spectroscopic radial velocity method measures the doppler velocity signatures of individual stars at multiple epochs, which is a very time consuming procedure. The photometric transit method can yield the light curves of thousands of stars simultaneously. More importantly, the photometric transit method provides information on planetary radius and the inclination of the planetary orbit relative to the line of sight, not possible from radial velocity detections. In addition, a wide array of studies are possible for transiting exoplanets, which cannot be done with non-transiting systems, e.g. the study of planetary atmospheres \citep{Sing2009}, temperature, surface brightness \citep{Snellen2007, Snellen2010}, and the misalignment between the planetary orbit and the stellar spin \citep{Win2005}.

Ideally, to search for transit exoplanet requires high-quality, wide-field, long-baseline continuous time-series photometry. This kind of monitoring can be achieved effectively by the ambitious space-based programs such as \textit{CoRoT} \citep{Baglin2006} and \textit{Kepler} \citep{Borucki2010} or complicated longitude-distributed network programs such as HATNet \citep{Bakos2004} and HATSouth \citep{Bakos2013}. However, the circumpolar locations offer a potentially comparable alternative.

The circumpolar locations provide favorable conditions for a wide and diverse range of astronomical observations, including photometric transiting detections. Thanks to the extremely cold, calm atmosphere and thin turbulent surface boundary layer, as well as the absence of light and air pollution, we can obtain high quality photometric images in circumpolar locations \citep{Burton2010,steinbring2010,steinbring2012,steinbring2013}. Furthermore, the long polar nights offer an opportunity to obtain continuous photometric monitoring. As shown by a series of previous thorough and meticulous studies (cf. Pont \& Bouchy 2005; Crouzet et al. 2010; Daban et al. 2010; Law et al. 2013), it greatly increases the detectability of transiting exoplanets, particularly those with periods in excess of a few days. Additionally, decreased high-altitude turbulence will result in reduced scintillation noise that will lead to superior photometric precision \citep{Kenyon2006}. The significant photometric advantages of the polar regions have been proven and utilized by the observing facilities at different polar sites such as two AWCam \citep{Law2013} at Canadian High Arctic, SPOT \citep{Taylor1988} at the South Pole, small-IRAIT \citep{Tosti2006}, ASTEP-South \citep{Crouzet2010} and ASTEP-400 \citep{Daban2010} at Dome C.

Dome  A, located in the deep interior of  Antarctica, with the surface elevation $4,093\,\rm{m}$, is the highest astronomical site on the continent and is also one of the coldest places on Earth. In a study that considered the weather, the boundary layer, airglow, aurorae, precipitable water vapor, surface temperature, thermal sky emission, and the free atmosphere, \citet{Saunders2009} concluded that Dome A might be the best astronomical site on Earth.

\begin{figure}
\epsscale{0.85}
\plotone{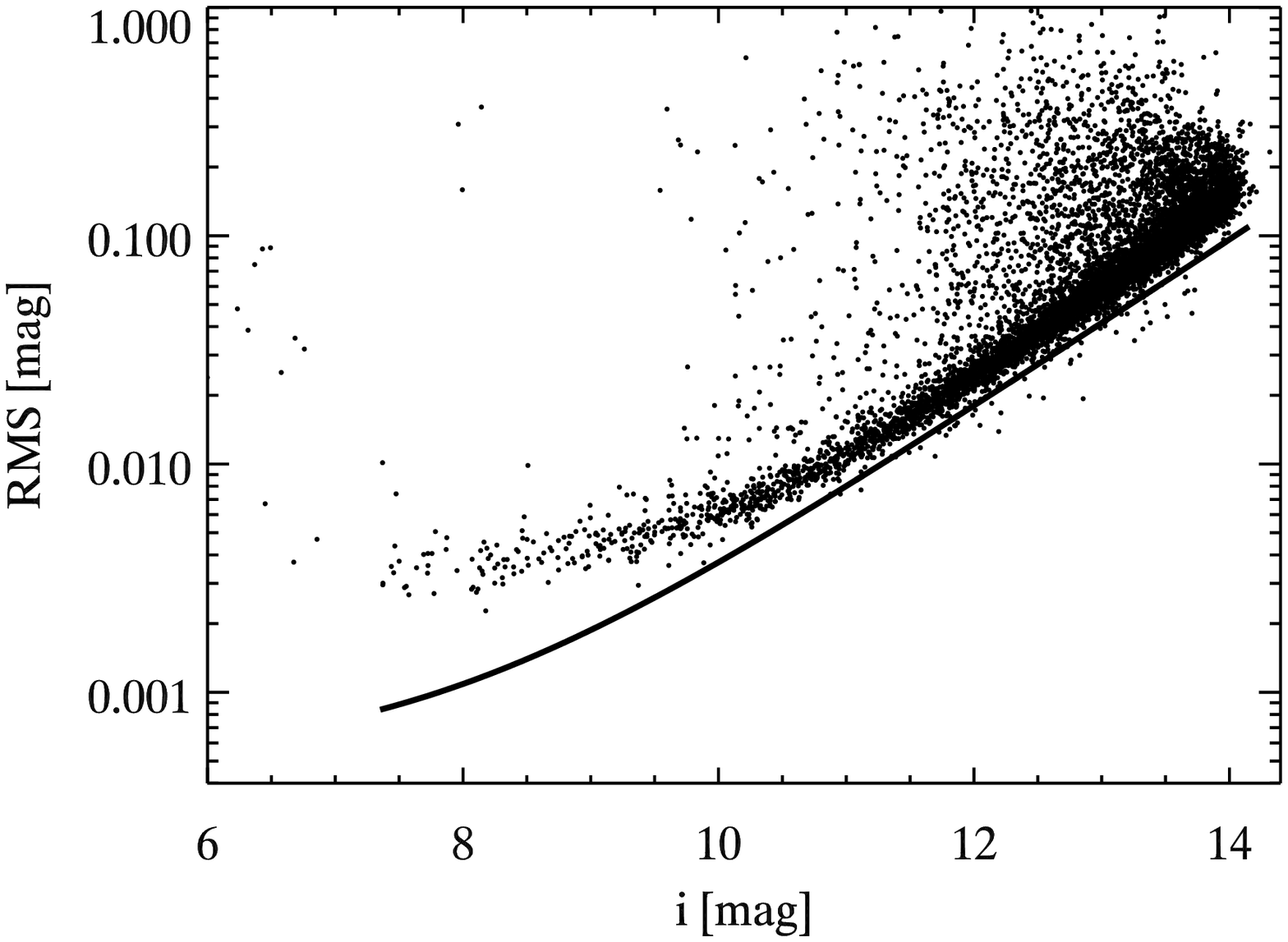}
\caption{The distribution of RMS values at 20-second cadence as a function of CSTAR $i$ magnitude. Each point represents a light curve. Photometric precision of resulting CSTAR light curve is typically $\sim 20\,\rm{mmag}$ at $i=12$, with $\sim 4\,\rm{mmag}$ achieved at $i=7.5$. We over-plotted the theoretic RMS as a function of magnitude, taking into account the photon and sky background noise as well as the scintillation noise.
\label{fig1}}
\end{figure}

In order to take the advantage of these remarkable observing conditions at the Dome A, the Chinese Small Telescope ARray (CSTAR) was established at Dome A in 2008 January. CSTAR undertook both site testing and science research tasks. In 2008, 291,911 qualified \textit{i}-band photometric images were acquired. Based on these data, the first version of photometric catalog has been released by \citet{Zhou2010a}, and updated three times \citep{Wang2012, Wang2013, Meng2013} to correct for various systematic errors. The resulting CSTAR photometric precision typically reaches $\sim 4\,\rm{mmag}$ at 20-second cadence at $i=7.5$, and is $\sim 20\,\rm{mmag}$ at $i=12$ (see Figure~\ref{fig1}), which is sufficient for the detection of giant transiting exoplanets around F, G, K, dwarf stars.

In this paper, we present ten exoplanet candidates to come from 10,690 high precision light curves selected from the CSTAR data of 2008 \citep{Wang2013}. From all these candidates four were found to be giants using spectroscopic follow-up. Since this is the first effort to find  exoplanets from these data, we describe the CSTAR instrument, observations, previous data reductions and the methods used for the transit searching in detail, as well as the procedures used to eliminate the false positives.

The layout of the paper is as follows. A brief description of the CSTAR instrument, observations and previous data reduction, as well as the photometric precision of the light curves, is presented in Section 2. In Section 3, we detail the techniques we used for transit detection and the robust procedures of data validation. The spectroscopic and radial velocity follow-up are briefly described in Section 4. We report the exoplanet candidates along with the detailed properties for each system in Section 5. Lastly, the work is summarized and prospects for future work are discussed in Section 6.
\section{Instrument, Observations and Previous Data Processing}

\subsection{Instrument}
CSTAR, as a part of PLATeau Observatory (PLATO) \citep{Lawrence2009, Yang2009}, is the first photometric instrument to enter operation at Dome A. Full details of CSTAR instrument can be found in \citet{Yuan2008} and \citet{Zhou2010b}. Here we summarize the features relevant to this work. The CSTAR facility consists of four static, co-aligned Schmidt-Cassegrain telescopes on a fixed mount with the same $4^\circ.5 \times 4^\circ.5$ Field of View (FOV) around the South Celestial Pole, each telescope housing a different filter in SDSS bands: \textit{r, g, i} and \textit{open}. Each telescope gives a $145\,\rm{mm}$ entrance pupil diameter (effective aperture of $100\,\rm{ mm}$) and is coupled to a $1\rm K \times 1 \rm K$ Andor DV 435 frame transfer CCD array which yields the plate-scale of $15\,\rm{arcsec}\,\rm{pixel}^{-1}$.

\subsection{Observations}
CSTAR was successfully shipped and deployed at Dome A in 2008 January and operated for the following four years. This work is based on the data obtained in 2008.
In the 2008 observing season (2008 March 4 to August 8), intermittent problems with the CSTAR computers and hard disks \citep{Yang2009} prevent us from obtaining useful data in the \textit{g, r} and \textit{open} band. Fortunately the \textit{i} band data were not affected, and observations were carried out for $1728\,\rm{hours}$ (291,911 qualified frames with 20-second exposure times) during the Antarctic polar nights, with only a few short interruptions due to cloudy weather \citep{Zou2010} or temporary instrument problems \citep{Yang2009}. These observations provide well-sampled light curves with a baseline of more than one hundred days. Additional details of the CSTAR observations in 2008 are presented in \citet{Zhou2010a}.

\subsection{Previous Data Reductions}
Reduction of the CSTAR data aim to produce millimagnitude photometric precision for the bright stars. A custom reduction pipeline was developed which is able to achieve this goal and is described in more detail in \citet{Zhou2010a} and \citet{Wang2012, Wang2013} as well as \citet{Meng2013}. Here we will only briefly review the main factors to be considered when reducing the wide-field data from CSTAR.

After preliminary reductions, aperture photometry was performed on the sources that were detected in the all calibrated images. Using 48 brightest local calibrators, the instrumental magnitudes were calibrated to the $i$ magnitudes of the stars in the USNO-B 1.0 catalog \citep{Monet2003}, which were derived from the UNSO-B 1.0 magnitudes according to the transformation between USNO-B 1.0 magnitude and SDSS $i$ magnitude given by \citet{Monet2003}. Finally, the first version of CSTAR catalog, detailed in \citet{Zhou2010a}, was released.

For transit searching, the photometric data were further refined by applying corrections for additional systematic errors, as briefly reviewed below.

Poor weather will lead to spatial variations in extinction across the
large CSTAR FOV ($4.5^{\circ} \times 4.5^{\circ}$). This spatially
uneven extinction can be modelled and corrected by comparing each frame to a master (median) frame. The more detailed procedures has been described in \citet{Wang2012}.

The residual of the flat-field correction results in spatially dependent errors, which show up as daily variations when the stars are centered on the different pixels in different exposure frames during their diurnal motion around the south celestial pole on the static CSTAR optical system. This kind of diurnal effect can be effectively corrected by specific differential photometry: comparing the target object to a bright reference star in the nearby diurnal path. For more details, see \citet{Wang2013}.

Since CSTAR is a static telescope and fixed to point at the South
Celestial Pole, star images move clockwise on the CCD due to diurnal
motion. Ghost images, located in symmetrical position of the CCD, move
counterclockwise. For that reason, ghost images move and contaminate the
photometry of stars. The significant contamination arising from the ghost images, detailed in \citet{Meng2013}, was also studied and corrected.

The resulting light curves typically achieve a photometric precision of $\sim 4\,\rm{mmag}$ at 20-second cadence for the brightest non-saturated stars ($i=7.5$), rising to $\sim 20\,\rm{mmag}$ at $i=12$. The distribution of RMS values as a function of \textit{i} magnitude is shown in Figure~\ref{fig1}. Each of points represents a 20-second sampled light curve with one-day observations. The abrupt upturn in variability at $i < 7.5$ signifies the onset of saturation, and our photometry is complete to a limiting magnitude of $i = 14$. For that reason, we use the $i$-band time-series data on the 10,690 point sources, restricted to $7.5 < i <14$ in our study, to detect transit events.

\section{Transit Detection}

\subsection{Transiting Searching Algorithm}
To search for planetary transits in the light curves, the BLS algorithm \citep{Kov2002} is applied to the data. The search is limited within $1.05-30.0\,\rm{days}$ periods range, with 4500 period steps, 1,500 phase bins, and fractional transit length from $q_{\rm{min}}=0.01$ to $q_{\rm{max}}=0.1$. The BLS spectra of CSTAR light curves generally display an increasing background power towards the lower frequency. This is caused by slight long-term systematic trends of the light curves \citep{Bakos2004}. To remove this effects from the BLS spectra, a fourth-order polynomial is fitted and then subtracted from the spectra. For the most significant residual peak which do not lie at a known alias, fit statistics and parameters of the box-fitting transit model are obtained and then used to provide a ranked list of the best candidates.

\subsection{Candidate Selection Criteria}
The systematic errors and true astrophysical variabilities, such as low-mass star, ``blended stellar binaries", and ``grazing stellar binaries", can mimic the true transit signals and result in a high false-positive rate. For this reason, it is imperative to distinguish false-positive signals from the true exoplanet candidates. This section describes the procedures of candidate inspection based on the techniques used in previous successful transit surveys, such as WASP \citep{Pollacco2006}, HATNet \citep{Bakos2004}, HATSouth \citep{Bakos2013}, \textit{CoRoT} \citep{Baglin2006}, Optical Gravitational Lensing Experiment (OGLE) \citep{Udalski2002}, \textit{Kepler} \citep{Borucki2010}, XO \citep{McCullough2005}, and Trans-Atlantic Exoplanet Survey (TrES) \citep{Alonso2004}.
\subsubsection{Stage 1: Pre Filter}
As described in section 2.3, a total of 10,690 stars with sufficiently high precision were selected from the CSTAR data set for transit searching. They are processed by the detection algorithm, yielding an output of fit statistics and parameters of the box-fitting transit model. The large number of stars make visual inspection of every light curve infeasible. So we require that a number of conditions should be satisfied before subjecting the candidates to visual inspection. To avoid missing any interesting candidates before visual inspection, the initial selection criteria are deliberately set relatively low. The thresholds for rejection are:
\begin{itemize}
  \item \textit{Photometric transit depth greater than 10 percent.} The fractional change in brightness of transit depth is essentially determined by the square of the ratio of the planet radius to the host star radius. Giant transiting planets typically have depths on the order of one percent. We set a relatively loose depth criteria (10 percent) to avoid loss of interesting objects. Although a $\rm{R=2\,R_J}$ planet will block out a quarter of the light of late-type stars (e.g. M0 V star), as \citet{Kane2008} pointed out, these kinds of detections from bright, wide-field surveys would be extremely rare.
  \item \textit{Frequencies with empty phases.} The incomplete phase coverage leads to aliasing and can often cause false-positive detection. We use a simple model to exclude frequencies with poor phase coverage. The folded light curve is split with the expected transit width. A frequency is considered systematic if the number of empty intervals is larger than 2.
  \item \textit{Period $<$ $1.05\,day$ or periods at a known alias} The BLS algorithm, similar to other pattern matching methods, suffers from aliasing effects originating from nearly periodic sampling \citep{Kov2002}. Therefore, it creates false frequency peaks at period associated with one sidereal day and uniform 20-second sampling interval. The BLS spectra clearly display such peaks, as well as some other commonly occurring frequencies associated with the remaining systematic errors. For that reason stars exhibiting these periodicities are excluded in order to minimize the number of aliases. We have also elected not to search for transits with periods less than $1.05\,\rm{day}$, due to the large number of false frequency peaks in that region.
\end{itemize}

Even these relatively low selection criteria remove more than 85 percent of the initial detections. Only 1,583 candidates pass and these are then visually inspected as set out in the next section.

\subsubsection{Stage 2: Visual Inspection}
Our visual inspection procedure is based upon that used for the successful WASP program as described in \citet{Clarkson2007}, \citet{Lister2007} and \citet{Street2007}.

During the visual inspection of the folded light curves in conjunction with the corresponding BLS spectra, surviving candidates are require to have:
\begin{itemize}
\item \textit{Plausible transit shape.}
Since the transit depth has been limited in the \textit{stage 1},
transit shape becomes the first important aspect in this stage. A
visible transit dip is a basic requirement for a candidate to be called ``transit candidate".
\item \textit{Flat out-of-transit light curve.}
The light curve before and after transit should be flat. Candidates are removed if they show the clear evidence of variability out of transit, including the secondary eclipse, ellipsoidal variation as well as realistic variability of other form.
\item \textit{Smoothly phase coverage.}
Although candidates were systematically removed in \textit{stage 1} if their frequencies are associated with gaps in the folded light curves, some with uneven distribution of data points in the folded light curves, which may not be effectively identified in the \textit{stage 1}, were deselected from further consideration by visual inspection.

This step is also used to discard light curves of poor quality.
\item \textit{Credible measured period.}
BLS spectra together with the folded light curves are inspected to confirm whether the clear period peaks are arising from secure transit signals or other variabilities.
\end{itemize}

As the visual inspection process is somewhat subjective, it was carried out independently by the two authors (Songhu Wang and Ming Yang). After a comparison of the analysis, this examination reduced the 1,583 candidates down to 208 transit-like candidates, which required further investigation.

\subsubsection{Stage 3: Statistical Filter}
The main purpose of this stage is to facilitate the further identification of the true planetary candidates from false-positive transit detections caused by systematic trends or true astrophysical variability. Candidates are passed forward if:
\begin{itemize}
  \item \textit{Signal-to-red noise $(S_{\rm{r}}) \ge 7.0$}. Contrary to the white-noise (uncorrelated-noise) assumptions, the errors on ground-based millimagnitude photometry are usually red (correlated) \citep{Pont2006}. In the CSTAR data, the uncertainty of the mean decreases more slowly than $n^{1/2}$, suggesting that red-noise is present. This can mimic transit signal with a time-scale similar to the duration of the true close-in planetary transit. So, $S_{\rm{r}}$, a simple and robust statistical parameter to assess the significance level of detected transit in the presence of red noise, is calculated for each light curve by
\begin{equation}
S_{\rm r} = \frac{d \sqrt{N_{\rm tr}}}{\sigma_{\rm r}},
\end{equation}
where $d$ is the best-fitting transit depth, $N_{\rm tr}$ is the number of transits observed, $\sigma_{\rm r}$ is the uncertainty of transit depth binned on the expected transit duration in the presence of red noise. The simplest method of assessing the level of red-noise ($\sigma_{\rm r}$) present in the data is to compute a sliding average of the out-of-transit data over the $n$ data points contained in a transit-length interval.
      This method is proposed by \citet{Pont2006} and has been successfully applied to the SuperWASP candidates \citep{Christian2006, Clarkson2007, Kane2008, Lister2007, Street2007}.
      The typical level of $\sigma_{\rm r}$ in the CSTAR data is of $2.1\,{\rm mmag}$. It is slightly lower when compared to $3\,\rm{mmag}$ for the OGLE \citep{Pont2006} and SuperWASP \citep{Smith2006}. For that reason, although there is no confirmed planetary transit and no simulation was performed for the $S_{\rm r}$ threshold in the CSTAR survey, to attempt to detect more transiting planets, it is reasonable to set our $S_{\rm r}$ threshold to the lower boundary of the typical range (7-9) of that given by \citet{Pont2006} based on the detailed simulation with $S_{\rm r}=3\,\rm{mmag}$. This threshold is also consistent with that used for the SuperWASP candidates \citep{Christian2006}.
  \item \textit{The transit to antitransit ratio $(\Delta \chi^2/\Delta \chi^2_-) \ge 1.5$}. The systematic variations and the stellar intrinsic variables with timescale similar to the planetary transit can give rise to false-positive transit detections. A light curve with a genuine transit will result in only a strong transit (dimming) detection and not a strong antitransit (brightening) detection. On the contrary, one could expect the strong correlated measurements caused by the systematics or the stellar intrinsic variables should produce both significant transit and antitransit detections. Consequently, $\Delta \chi^2/\Delta \chi^2_-$, measuring the ratio of improvements of best-fit transit to the improvements of the best-fit antitransit, is calculated for each light curve. This provides an estimate to which a detection has the expected properties of a credible transit signal rather than the properties of the systematics or intrinsic stellar sinusoidal variability \citep{Burke2006}.

  \item \textit{Signal to noise of the ellipsoidal variation $(S_{\rm{ellip}})<5.0$.} Blended systems, gazing eclipsing binaries and eclipsing systems with a planet-sized star (e.g. brown dwarf) are the most common astrophysical imposters that mimic a transiting planet signal. It can be very difficult to distinguish these systems from genuine transiting planets using the properties of the transit event itself (e.g. shape, depth, etc). Nevertheless, evidence of ellipsoidal variability, due to tidal distortions and gravity brightening, can be used to remove from the remaining candidates which have massive, and therefore not planetary companions. The method, proposed by \citet{sirko2003}, was successfully applied to the OGLE \citep{Udalski2002} and the WASP \citep{Pollacco2006} candidates.

  \item \textit{No statistical differences between odd and even transits}. A blended or grazing eclipsing binary system can produce a shallow dip similar to an exoplanet transit. A true exoplanet would ideally lead to the evenly spaced transits with the same depths. In contrast, the depths of primary and secondary eclipses of a blended or grazing eclipsing binaries are generally different due to the difference in size and temperature of the two components. In addition, the primary and secondary eclipse are usually unevenly spaced in the time series since the orbit of binaries is generally eccentric \citep{Wu2010}. We use the significant level of the consistency in transit depth ($P_\delta$) and epoch ($P_t$), as detailed in \citet{Wu2010}, to assess whether the odd and even transits are drawn from the same population. The smaller this statistic, the more likely the event is an astrophysical false positive. The significant level ($P_\delta$ or $P_t$) of 0.05 or less denotes the transit signal is unlikely to be caused by a transiting planet.
  \item \textit{No aperture blends}. Blended eclipsing binary systems are some of the most common imposters identified as the transiting planets in wide-field transit surveys such as CSTAR. The large plate-scale of CSTAR makes it likely that there will be more than one bright object within a single CSTAR pixel ($15\, \rm{arcsec}$) or the applied photometric aperture ($\rm{radius}=45\, \rm{arcsec}$) of the CSTAR photometry. This can lead to a dilution of depth of a stellar eclipsing binary, making it appear similar to a transiting exoplanet. If the angular separation of the blend is less than or comparable to the pixel scale of CSTAR, we cannot eliminate the false positive arising from blended eclipsing in this step, however, imposters arising from the wider blends can be eliminated here: The candidates are eliminated if the center of a brighter object is present within 45-arcsec aperture.

      In addition, for some candidates, aperture photometry is subject to contamination by nearby bright objects (just outside the photometric aperture). The detected transit-like shallow dip could be due to the nearby object with a deep eclipse. These spurious candidates are rejected by comparing their light curves to those of nearby objects.

\end{itemize}

     We note that to avoid missing some interesting systems, some candidates with parameters just outside these thresholds have also been carried forward to the next stage. We find just ten candidates of the initial 208 candidates pass through these statistical filters.

\subsubsection{Stage 4: Additional System Information}
The ten candidates which pass through the third stage are analyzed in the following manner:
\begin{itemize}
  \item \textit{Stellar information}. To estimate the radius of the transiting candidate, the radius of the host star must be determined. The color indices, derived from Tycho-2 $B-V$ \citep{Hog2000}, are used to estimate the spectral type and radius of the host stars based on the data from \citet{Cox2000}, assuming the host stars to be main sequence.

       Using the besancon model \citep{Robin2003} we estimate that $40\%$ of the stars in our FOV between $i=7.5-12$ are giants, for which the detected transit signals would then due to other stars, not planets. Taking \citet{brown2003} as a guide, the 2MASS $J-K$ colors \citep{cutri2003} can act as a rough indicator of the luminosity class of the target. Candidates with $J-K>0.5$ are flagged as potentially giants.
  \item \textit{Refined transit parameters}. The remaining transit light curves are modelled using the jktebop code \citep{Southworth2004}. The refined parameters of these system, such as period, epoch, particularly the planetary radius ($R_{\rm p}$), are obtained from these modeling results together with the derived host stellar radius ($R_*$). Although gas giant planets, brown dwarfs and white dwarfs can all have similar radii, we regard CSTAR candidates with estimated radii less than $2\,\rm{R_{Jup}}$ as realistic candidates.
  \item \textit{The ratio of the theoretical duration and the observed duration ($\eta$)}. For each candidate, we provide the ratio of the theoretical duration and the observed duration ($\eta$), which is introduced for the OGLE candidates by \citet{Tingley2005} and then has been successfully applied in the WASP candidates. $\eta$ of strong exoplanet candidate is expected to close to 1.
\end{itemize}

    The analysis set out in this section was only to provide additional information to remaining system but we did not use it to cull any candidates.

\section{Follow-Up Observations}
In this section we describe the follow-up spectroscopy that we have undertaken to help identify two common sources of false positives in transit surveys: eclipses around giant host stars and eclipsing binaries.

\subsection{Spectral Typing Follow-Up}

If a candidate host star is a giant, then its large stellar radius means that the transit event see in the discovery data cannot be due to a transiting exoplanet. We therefore spectral typed each of the 10 candidates to check for giant hosts. On the night of 2013 September 9 we took a single spectrum of each candidate with the Wide Field Spectrograph \citep{Dopita2007} on the Australian National University (ANU) $2.3\,\rm{m}$ telescope. Spectra we taken using the B3000 grating which results in a resolution of R=3000 and a wavelength range of 3500 to 6000~\AA. Spectra were reduced and flux calibrated in accordance with the methodology set out in \citet{Bayliss2013}. The spectra were compared to a grid of template spectra from the MARCS models \citep{Gustafsson2008}. The candidates \textit{CSTAR J021535.71-871122.5}, \textit{CSTAR J014026.01-873057.1}, \textit{CSTAR J203905.43-872328.2} and \textit{CSTAR J231620.78-871626.8} all showed \ensuremath{\log{g}}$<3.1$, indicating that they are giants and can be ruled out as candidates. The six remaining candidates are dwarfs and we therefore continued with multiple epoch RV measurements for these candidates to check for high-amplitude RV variations indicative of eclipsing binaries.

\subsection{Radial Velocity Follow-Up}

For the six dwarf candidates we obtained multi-epoch radial velocity measurements using WiFeS with the R7000 grating. Details on the technique for obtaining radial velocity measurements on WiFeS are set out in \citet{Bayliss2013}. On nights spanning 2013 September 20-25 we took between 3 and 5 RV measurements for each six candidates spanning a range of phases for each candidate. None of the candidates showed any RV variation beyond the intrinsic measurement scatter of $2\,\rm{km\,s^{-1}}$, indicating that none of these unblended eclipsing binaries.  All six therefore remain as good candidates for future high resolution radial velocity follow-up and/or photometric follow-up.

\section{Result and Discussion}

This section we present the ten CSTAR candidates in detail and discuss the follow-up observations we have made for each candidate.

\subsection{Result of transit search}
The candidate selection process result in ten promising exoplanet candidates, four of them were found to be giants using spectroscopic follow-up. Med-resolution radial velocity showed none of the remaining six candidates have radial velocity variation great than $2\,\rm{km\,s^{-1}}$. All of these candidates are listed in Table~\ref{table1}, along with the detailed information of them. The candidate ID is of the form `CSTAR J$hhmmss.ss-ddmmss.s$', with the position coordinates based on Tycho (J2000.0) position \citep{Hog1998}.

\begin{figure}
\epsscale{0.85}
\plotone{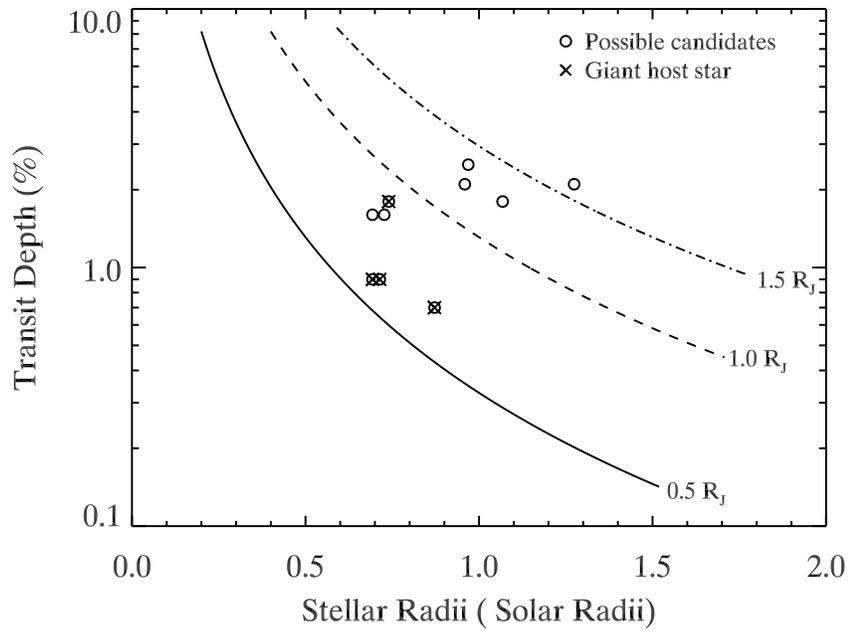}
\caption{The transit depth plotted as a function of stellar radius. Over-plotted are curves showing the expected transit depth for planets with radii of 0.5, 1.0 and $1.5\,\rm{R_J}$ assuming centrally crossing transit ($i=90 \degr$).
\label{fig2}}
\end{figure}

In Figure~\ref{fig2} we plot the theoretical curves of transit depth produced by planets of 0.5, 1.0 and 1.5 Jupiter radii as a function of host star radius assuming central crossing transit ($i=90\degr$). All of candidates are shown as open circles. Those with giant host stars are over-plotted as crosses. It can be seen all the six remaining candidates have reasonable planetary radii between 0.5 and $1.6\,\rm{R_J}$.

\subsection{Discussion of Candidates}
In this section we provide a detailed description of each of ten candidates. In addition, and for completeness, we also discuss the system `CSTAR J183056.78-884317.0', an eclipsing binaries with a light curve that is similar to a transiting exoplanet light curve and which has been previously identified by other groups. The details are summarized in Table~\ref{table1}. The binned phase-folded light curves of these candidates along with their respective BLS periodograms are shown in Figures 5 to 14.

\begin{figure}
\epsscale{0.85}
\plotone{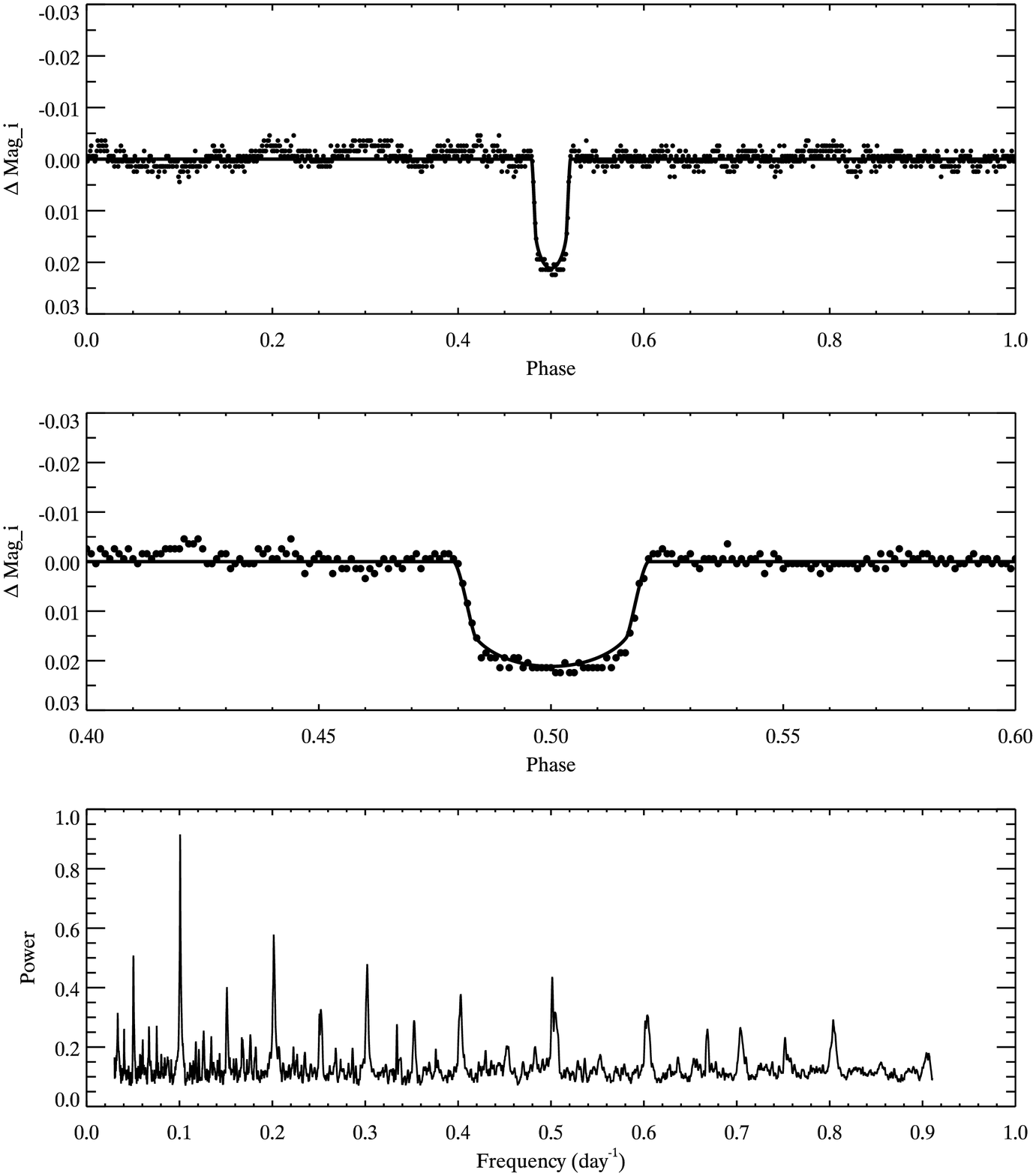}
\caption{The full (top panel) and zoom-in (middle panel) binned phase-folded ($p=9.924\,\rm{d}$) light curve (filled circles) along with the normalized and detrended BLS periodogram (bottom panel) of CSTAR J183056.78-884317.0 ($i=9.84$). The solid line in the top and middle panel show the best-fit transit model (JKTEBOP). For clarity, the phased light curve was binned into 1,000 bins. The binned light curve is shown for visualization only and was not used in our analysis.
\label{fig3}}
\end{figure}

\begin{figure}
\epsscale{0.85}
\plotone{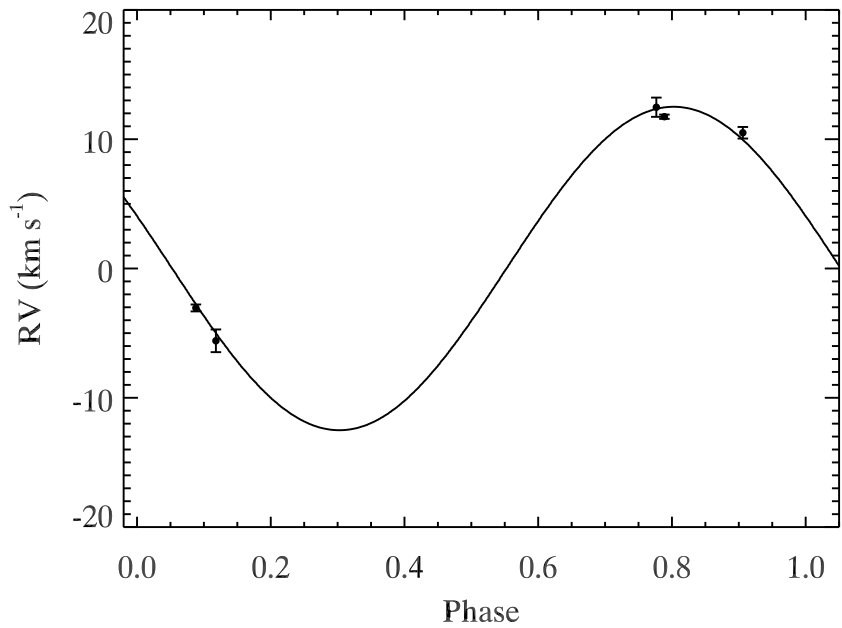}
\caption{Radial Velocity measurements (filled circles) for CSTAR J183056.78-884317.0 from the WiFeS and echelle instruments on ANU $2.3\,\rm m$ telescope, together with $e=0$ fit model (solid curve). The semi-amplitude of the best fit $e=0$ orbit gives $K=12\,\rm{km\,s^{-1}}$, indicating this is an eclipsing binaries system. This is consistent with our results derived from analysis of the transit duration and ellipsoidal variation.
\label{fig4}}
\end{figure}

\begin{figure}
\epsscale{0.85}
\plotone{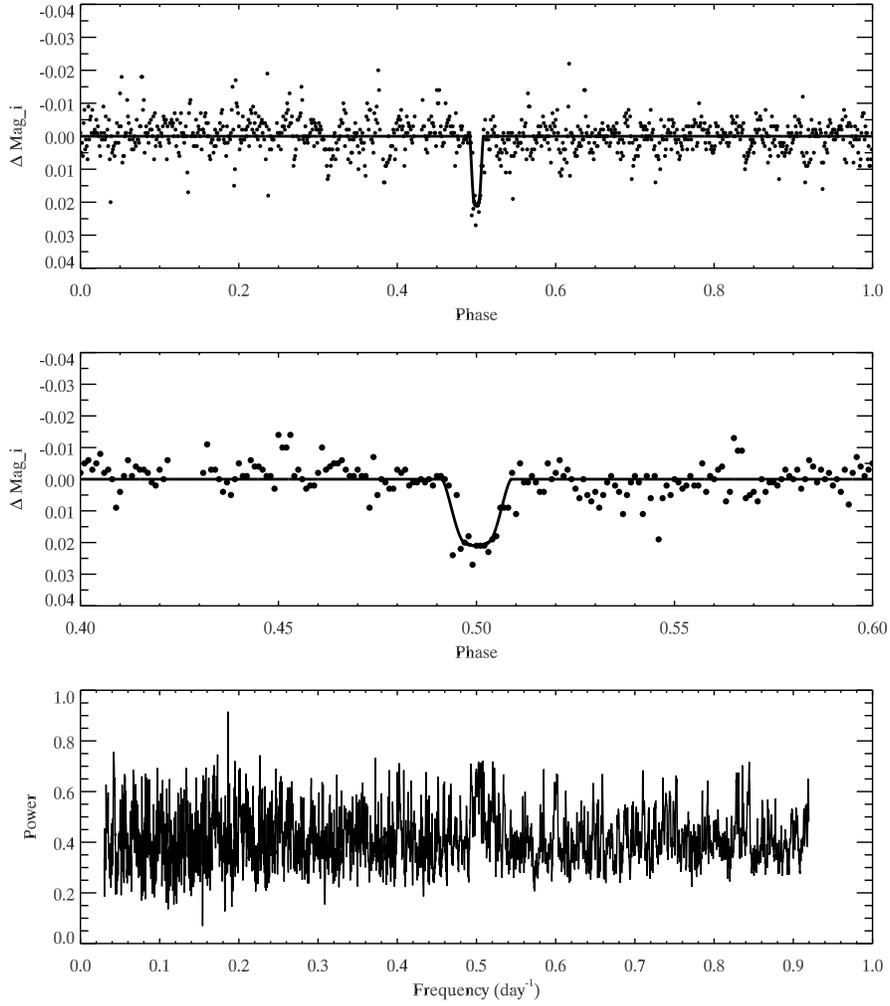}
\caption{Same as Figure~\ref{fig3}, but phase folded for $p=5.371\,\rm{d}$ for CSTAR J001238.65-871811.0 ($i=10.59$).
\label{fig5}}
\end{figure}

\begin{figure}
\epsscale{0.85}
\plotone{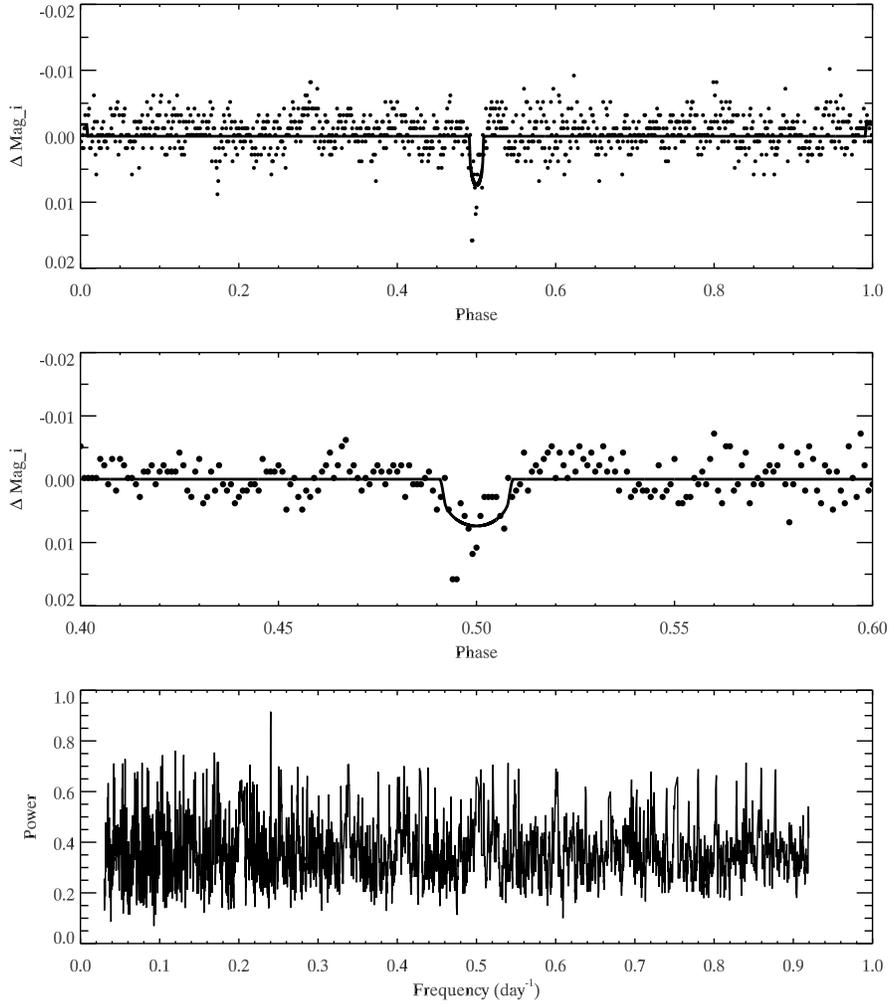}
\caption{Same as Figure~\ref{fig3}, but phase folded for $p=4.164\,\rm{d}$ for CSTAR J014026.01-873057.1 ($i=10.26$).
\label{fig6}}
\end{figure}

\begin{figure}
\epsscale{0.85}
\plotone{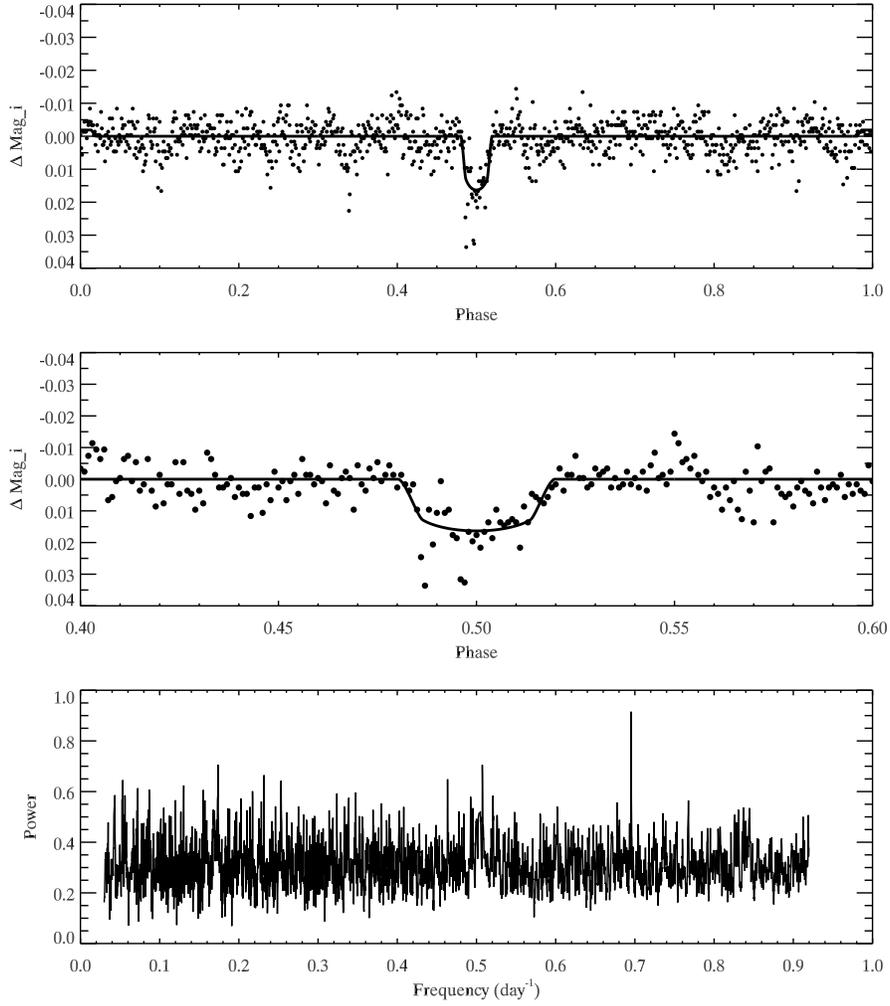}
\caption{Same as Figure~\ref{fig3}, but phase folded for $p=1.438\,\rm{d}$ for CSTAR J021535.71-871122.5 ($i=10.69$).
\label{fig7}}
\end{figure}

\begin{figure}
\epsscale{0.85}
\plotone{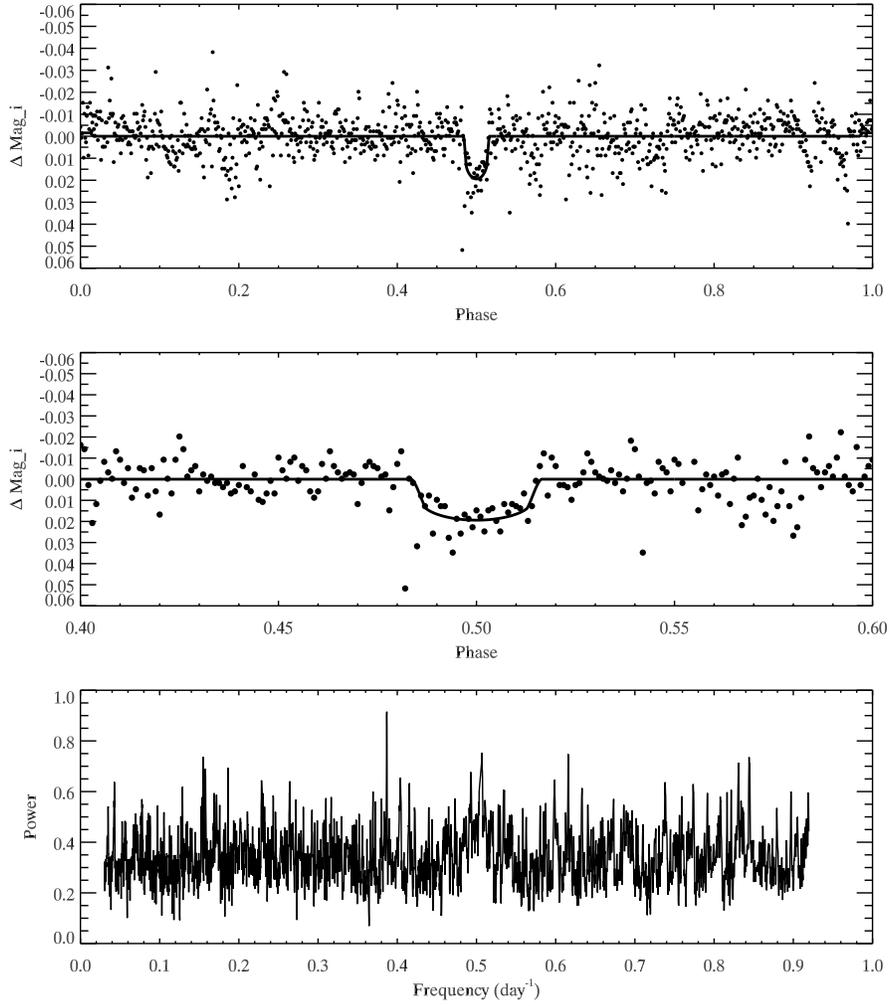}
\caption{Same as Figure~\ref{fig3}, but phase folded for $p=2.586\,\rm{d}$ for CSTAR J022810.02-871521.3 ($i=10.62$).
\label{fig8}}
\end{figure}

\begin{figure}
\epsscale{0.85}
\plotone{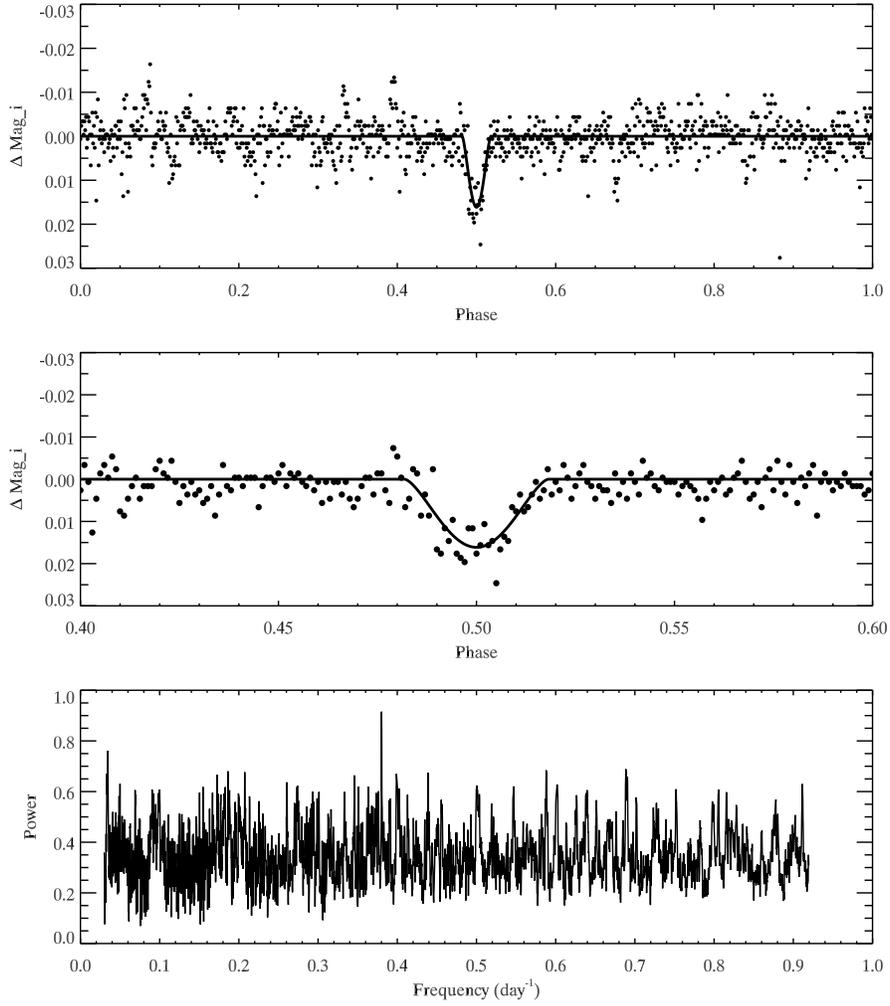}
\caption{Same as Figure~\ref{fig3}, but phase folded for $p=2.630\,\rm{d}$ for CSTAR J075108.62-871131.3 ($i=10.41$).
\label{fig9}}
\end{figure}

\begin{figure}
\epsscale{0.85}
\plotone{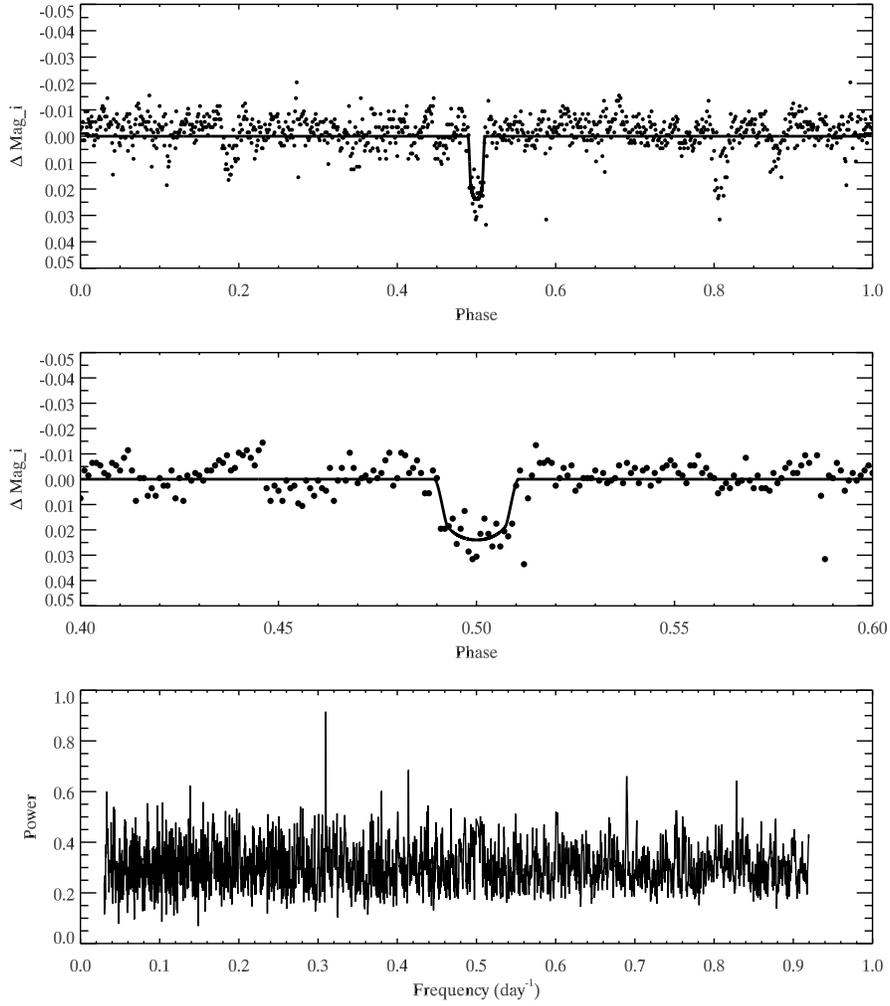}
\caption{Same as Figure~\ref{fig3}, but phase folded for $p=3.228\,\rm{d}$ for CSTAR J110005.67-871200.4 ($i=10.84$).
\label{fig10}}
\end{figure}

\begin{figure}
\epsscale{0.85}
\plotone{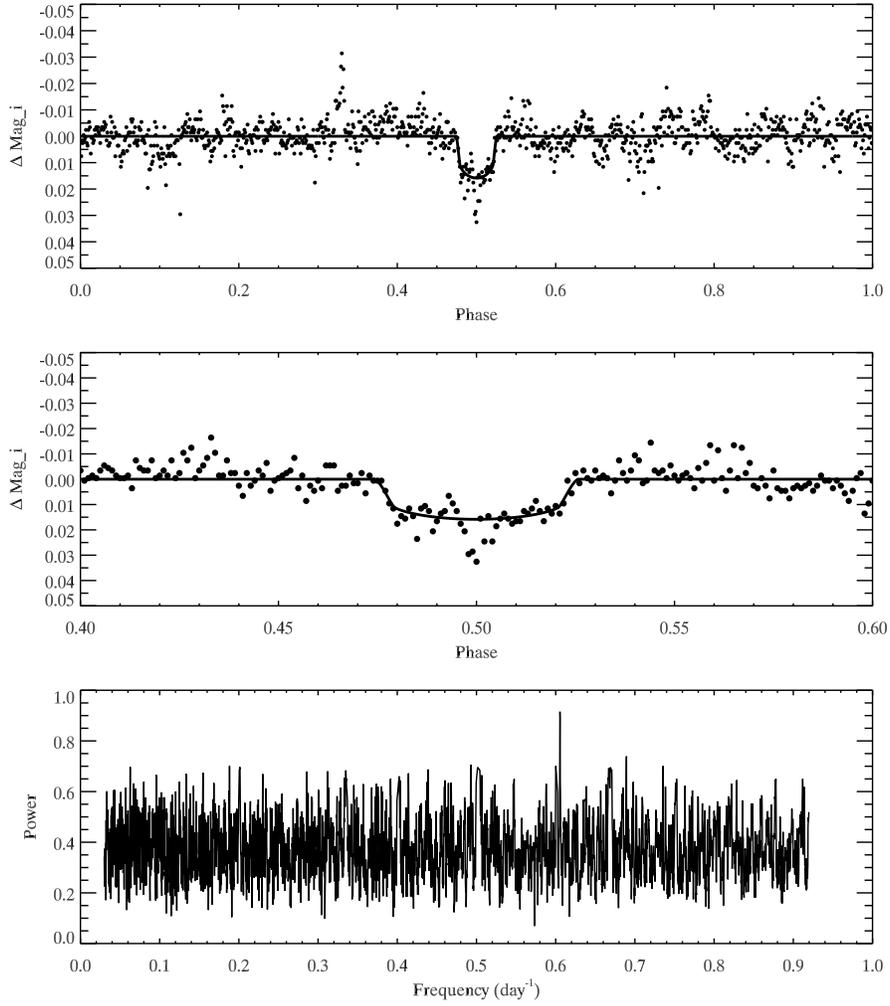}
\caption{Same as Figure~\ref{fig3}, but phase folded for $p=1.652\,\rm{d}$ for CSTAR J113310.22-865758.3 ($i=9.97$).
\label{fig11}}
\end{figure}

\begin{figure}
\epsscale{0.85}
\plotone{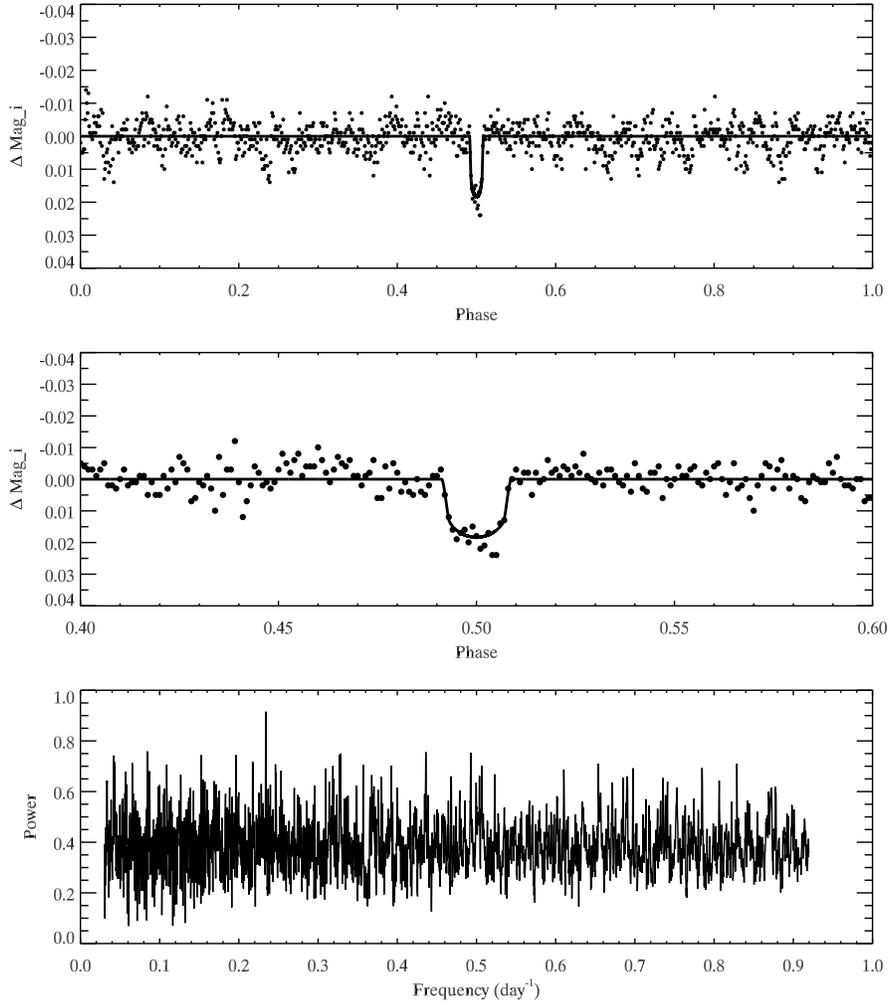}
\caption{Same as Figure~\ref{fig3}, but phase folded for $p=4.273\,\rm{d}$ for CSTAR J132821.71-870903.3 ($i=10.41$).
\label{fig12}}
\end{figure}

\begin{figure}
\epsscale{0.85}
\plotone{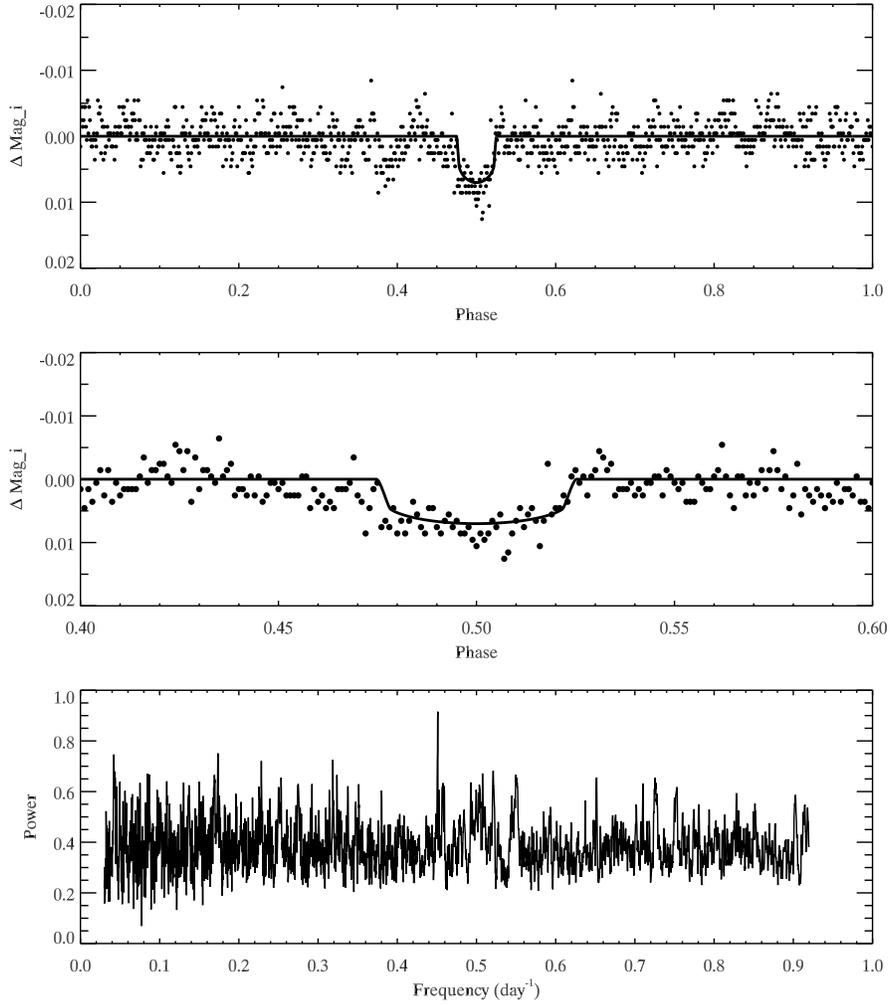}
\caption{Same as Figure~\ref{fig3}, but phase folded for $p=2.216\,\rm{d}$ for CSTAR J203905.43-872328.2 ($i=10.35$).
\label{fig13}}
\end{figure}

\begin{itemize}

  \item \textit{CSTAR J183056.78-884317.0}

  As shown in Figure~\ref{fig3}, this system exhibits a classic, flat-bottomed transit signature in the binned folded light curve of this bright ($i=9.84$) star and there is a strong periodic peak at $9.93\,\rm{d}$ from 13 detected transits. However, a relatively marked ellipsoidal variation ($S/N_{\rm{ellip}}=5.87$) together with a long duration ($\sim 10\,\rm{h}$) and high value of $\eta$ (2.03), suggest that it more likely to be an eclipsing binary.
  This object is also identified by the ASTEP team \citep{Crouzet2010} and another CSTAR analysis team \citep{Wang2011}. To verify our analysis results, the spectroscopic observations are applied to the object using both low-resolution Wide Field Spectrograph \citet{Dopita2007} and higher-resolution echelle on the ANU $2.3\,\rm m$ telescope. The results from five observations are presented in Figure~\ref{fig4} and show a radial velocity semi-amplitude of $K=12\,\rm{km\,s^{-1}}$, indicating that the candidate is an eclipsing binary. The ASTEP identification of this candidate is detailed in \citet{Crouzet2013}.

  \item \textit{CSTAR J001238.65-871811.0}

  This candidate has 24 transits with two percent depth and has the longest period ($5.37\,\rm d$) of the ten candidates.
  The companion radius of $0.96\, \rm{R_J}$ is supported by a slightly low but acceptable value of $\eta$ (0.65). As all parameters of this candidate easily pass the transit-sift threshold, it is worth high-priority follow-up, although there is a relatively large scatter in the light curve and periodogram (Figure~\ref{fig5}).

  \item \textit{CSTAR J014026.01-873057.1}

  As show in Figure~\ref{fig6}, the object displays a relatively shallow (0.9 percent) transit in an otherwise flat, if noisy, folded light curve with a well-defined period of $4.16\, \rm d$. The Tycho-2 color $(B-V=1.5)$ suggests a M4 primary with $0.71\,\rm{R_\odot}$, leading to a rather small planetary radius of $0.52\, \rm{R_J}$ and a reasonable $\eta=0.71$ if it was a dwarf. However, the very red color of the host star ($J-K=0.67$) suggested it was more likely to be a giant \citep{brown2003} and this was confirmed by our spectroscopic follow-up which gave $\rm{log}(g)=0.6$.

  \item \textit{CSTAR J021535.71-871122.5}

 Although there is some scatter in the light curve over the transit (Figure~\ref{fig7}), there is a strong peak in the periodogram. The observed short period ($1.438\,\rm d$) may place this candidate a very hot Jupiter. The exceptional high $\bigtriangleup x^2$/$\bigtriangleup x^2_{-}$ (2.69) and $S_{\rm{r}}$ (12.10) together with low $S/N_{\rm{ellip}}$ (0.48) plus well agreed odd- and even-transits make this seem to be a strong candidate. However, the infrared color of the host star ($J-K=0.80$) suggests this object may be a giant and this was confirmed by our spectroscopic follow-up ($\rm{log}(g)=3.3$).

  \item \textit{CSTAR J022810.02-871521.3}

  The object displays a transit with strong period ($2.586\,\rm d$) in an otherwise flat, if noisy, folded light curve (Figure~\ref{fig8}). The F-type primary star implies a $1.55\,\rm{R_{Jup}}$ companion (the largest companion of the ten candidates) and an acceptable $\eta$ (0.61). These factors together with the high $\bigtriangleup x^2$/$\bigtriangleup x^2_{-}$ (2.63) and low $S/N_{\rm{ellip}}$ (0.65) make this target a good candidates.

  \item \textit{CSTAR J075108.62-871131.3}

  This candidate displays a clear transit-like dip in the folded light curve (Figure~\ref{fig9}) and well meet all of the selection criteria. The low $S/N_{\rm{ellip}}$ (0.75) plus the high $S_{\rm{r}}$ (8.6) as well as  $\eta \sim 1$ make this a strong candidate. Although the very red color of the host star ($J-K=0.95$) suggests it may have been a giant, our spectroscopic follow-up ($\rm{log}(g)=4.5$) suggests it more likely to be a dwarf.

  \item \textit{CSTAR J110005.67-871200.4}

   As shown in Figure~\ref{fig10}, the transit in this candidate is obvious and there is a strong peak ($3.23\, \rm d$) in the periodogram. The high $S_{\rm{r}}$ (10.6) and $\bigtriangleup x^2$/$\bigtriangleup x^2_{-}$ (2.02) indicate the transit is not due to systematics. The $S/N_{\rm{ellip}}$ is low at 1.2 and the light curve is flat outside of transit.
   The estimate of the host radius and transit depth indicate a companion with moderate radius ($1.34\,\rm{R_{Jup}}$) and an acceptable, if a bit low, $\eta$ (0.55). The combination of these factors makes this candidate a high-priority target.

  \item \textit{CSTAR J113310.22-865758.3}

  This candidate displays a prototypical transit of one and half percent depth over an otherwise flat, if a little bit noisy, folded light curve (Figure~\ref{fig11}). The strong peak ($1.65\,\rm d$) in the periodogram together with low ellipsoidal variation ($S/N_{\rm{ellip}}=2.17$) as well as a reasonable $\eta =1.03$ indicated this brightest candidate ($i=9.97$) a good exoplanet candidate.

  \item \textit{CSTAR J132821.71-870903.3}

  The object clearly shows a `U'-shaped dip in an otherwise flat light curve (Figure~\ref{fig12}). This candidate has a relatively long period of $4.27\, \rm d$. We derive a reasonable radius ($1.26\,\rm{R_{Jup}}$) of the companion for its G0 spectral type. However, an acceptable, but relatively low $\eta$ (0.53) together with a slightly difference between odd-and even transit depth make this object a lower priority candidate.

  \item \textit{CSTAR J203905.43-872328.2}

   This object displays a very shallow ($\sim 0.007\, \rm mag$) but clear flat-bottom dip with a flat out of transit light curve (Figure~\ref{fig13}) which shows no signs of ellipsoidal variation ($S/N_{\rm{ellip}}=0.53$). There is a strong peak ($2.22\,\rm d$) in periodogram. The predicted relatively small companion radius of $0.64\,\rm{R_{Jup}}$ is slightly tempered by $\eta =1.15$. The relatively red 2MASS $J-K$ color (0.68) suggests a possible giant host star and it was confirmed by our spectroscopic follow-up which gave $\rm {log}(g)=1.5$.

  \item \textit{CSTAR J231620.78-871626.8}

  While noisy, this folded light curve (Figure~\ref{fig14}) exhibits a shallow transit. The strongest peak in the periodogram corresponds to $1.41\,\rm d$ which is the shortest companion of the final candidates. The derived radius ($0.69\,\rm{R_{Jup}}$) of companion are relatively small but the calculated transit duration is close to the observed one ($\eta =0.94$). However the relatively red color ($J-K=0.81$) suggested this object may be a giant and this was confirmed by our spectroscopic follow-up ($\rm {log}(g)=1.5$). We also note that the relatively low $S_{\rm{r}}\,(6.7)$ together with a slightly difference between odd- and even-transit depth indicated this candidate may have been a false positive.

\end{itemize}

\subsection{Discussion of Further Follow-up Observations}
The transit method has proven to be an excellent way of finding exoplanets, however final confirmation and determination of the planetary mass and radius requires high precision photometry and radial velocity follow up. Such observations of the candidates in our list are being performed by our colleagues at Australia now.

\section{Conclusion}
In 2008, more than 100 days of observations for a $20\,\rm{deg}^2$ field centered at the South Celestial Pole with the Antarctic CSTAR telescope provided high-precision, long-baseline light curves of 10,690 stars with a cadence of 20 seconds.

From this data set we found ten bright exoplanet candidates with short period. Subsequent spectral follow-up showed that four of these were giants, leaving six candidates. Med-resolution radial velocity showed none of the six candidates have radial velocity variation great than $2\,\rm{km\,s^{-1}}$. These detections have enriched the relatively limited optical astronomy fruit in Antarctica and indirectly reflects the favorable quality of Dome A for continuous photometric observations.

However, the real strength of CSTAR will be realized when the 2008 data are combined with the multi-color observations of following years. We expect to find many more candidates, especially those with longer periods and small radii, as a result of longer baseline along with higher signal to noise ratio.

The photometric data, including all of the CSTAR catalog and the light curves, are a valuable data set for the study of variable stars as well as hunting for transit exoplanets.

\acknowledgments
We thank the anonymous referee for the suggestions to improve the manuscript. This research is supported by the National Basic Research Program of China (No. 2013CB834900, 2014CB845704, 2013CB834902, 2014CB845702), the National Natural Science Foundation of China under grant Nos. 11333002, 11073032, 11003010, 10925313, 11373033, 11373035, 11203034, 11203031, the Strategic Priority Research Program-The Emergence of Cosmological Structures of the Chinese Academy of Sciences (Grant No. XDB09000000), 985 project of Nanjing University and Superiority Discipline Construction Project of Jiangsu Province, fund of Astronomy of the National Nature Science Foundation of China and the Chinese Academy of Science under Grants U1231113, the Natural Science Foundation for the Youth of Jiangsu Province (NO. BK20130547).





\clearpage


\clearpage

\begin{figure}
\epsscale{0.85}
\plotone{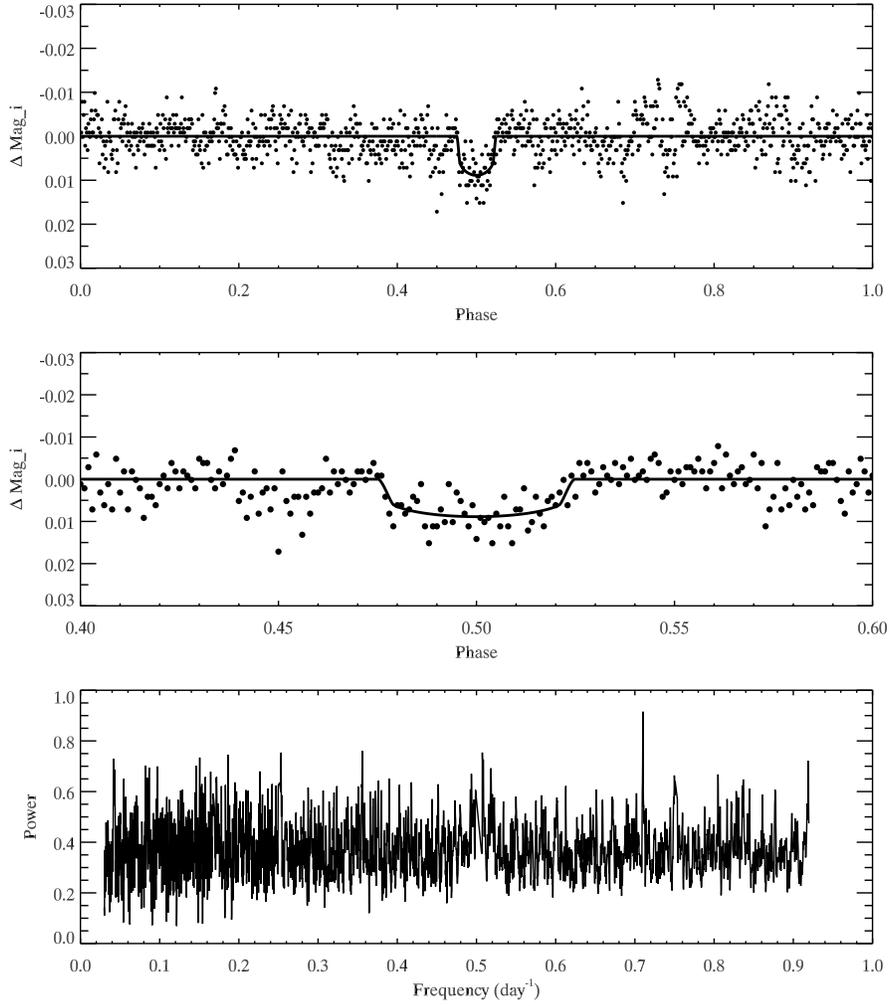}
\caption{Same as Figure~\ref{fig3}, but phase folded for $p=1.408\,\rm{d}$ for CSTAR J231620.78-871626.8 ($i=10.76$).
\label{fig14}}
\end{figure}


\setlength{\tabcolsep}{1.1pt}
\clearpage

\begin{deluxetable}{cccccccccccccccccc}
\rotate
\tabletypesize{\scriptsize}
\tablewidth{0pt}
\tablecaption{Summary of CSTAR exoplanet transit candidates}
\tablehead{
\colhead{CSTAR ID}   &  \colhead{Epoch}           &  \colhead{\it{i}}      &
\colhead{Period}     &  \colhead{Duration}        &  \colhead{Depth}       &
\colhead{$R_*$}      &  \colhead{$R_p$ }          &  \colhead{$B-V$}       &
\colhead{$J-K$}      &  \colhead{$\rm{T_{eff}}$}  &  \colhead{log(g)}      &
\colhead{$S_p$}       &
\colhead{$\bigtriangleup x^2$/$\bigtriangleup x^2_{-}$}                    &
\colhead{$S/N_{\rm{ellip}}$}                                               &
\colhead{$S_{\rm{r}}$}                                                   &
\colhead{$\eta$}     &  \colhead{$P_{\delta}\mid P_t$}    \\

\colhead{CSTAR J+}          &  \colhead{(2454500.0 +)}    &   \colhead{(mag)}      &
\colhead{(d)}               &  \colhead{(h)}              &   \colhead{(mag)}      &
\colhead{$(\rm{R_\odot})$}  &  \colhead{$(\rm{R_{Jup}})$} &   \colhead{(mag)}      &
\colhead{(mag)}             &  \colhead{K}                 &   \colhead{}           &
\colhead{}           &
\colhead{}                  &  \colhead{}                 &   \colhead{}           &
\colhead{}                  &  \colhead{}
}
\startdata
183056.78-884317.0 & 53.69665 &  9.84 & 9.924 & 10.004 & 0.021 & 1.214 & 1.531 & 0.48 & 0.31 & --- & --- & F5 & 4.23 & 5.87 & 22.32 & ~2.03~  & $0.42\mid 0.38$\\
\\
001238.65-871811.0 & 48.80221 & 10.59 & 5.371 &  2.269 & 0.021 & 0.959 & 1.356 & 0.69 & 0.43 & 5900 & 4.9 & G5 & 3.53 & 0.28 &  8.78 & 0.65 & $0.66\mid 0.74$\\
\\
014026.01-873057.1 & 46.69858 & 10.26 & 4.164 & 1.847 & 0.009 & 0.714 & 0.519 & 1.54 & 0.67 & 4800 & 0.6 & Giant & 1.48 & 0.26 & 10.37 & 0.71 & $ 0.15\mid 0.44$ \\
\\
021535.71-871122.5 & 46.50898 & 10.69 & 1.438 & 1.360 & 0.018 & 0.740 & 0.862 & 1.65 & 0.80 & 4600 & 3.3 & Giant & 2.69 & 0.45 & 12.10 & 0.71 & $ 0.48\mid 0.23 $\\
\\
022810.02-871521.3 & 50.90359 & 10.62 & 2.586 & 2.048 & 0.021 & 1.274 & 1.547 & 0.44 & 0.36 & 6100 & 3.5 & F5 & 2.63 & 0.65 & 7.11 & 0.61 & $ 0.64\mid 0.11 $\\
\\
075108.62-871131.3 & 47.59870 & 10.41 & 2.630 &  2.298 & 0.016 & 0.693 & 0.742 & 1.24 & 0.95 & 4800 & 4.5 & K7 & 1.52 & 0.75 &  8.60 & 1.02 & $0.17\mid 0.42$\\
\\
110005.67-871200.4 & 47.11239 & 10.84 & 3.228 & 1.633 & 0.025 & 0.969 & 1.335 & 0.68 & 0.33 & 6300 & 3.9 & G5 & 2.02 & 1.19 & 10.60 & 0.55 & $ 0.07\mid 0.62 $ \\
\\
113310.22-865758.3 & 47.14206 &  9.97 & 1.652 &  2.045 & 0.016 & 0.727 & 0.794 & 1.06 & 0.60 & 4900 & 5.0 & K4 & 1.63 & 1.72 &  6.96 & 1.03 & $0.45\mid 0.40$ \\
\\
132821.71-870903.3 & 46.53672 & 10.41 & 4.273 &  1.797 & 0.018 & 1.068 & 1.255 & 0.59 & 0.41 & 6000 & 4.5 & G0 & 1.62 & 2.17 &  7.05 & 0.53 & $0.01\mid 0.20$ \\
\\
203905.43-872328.2 & 47.21003 & 10.35 & 2.216 &  2.691 & 0.007 & 0.872 & 0.636 & 0.79 & 0.68 & 4800 & 1.5 & Giant & 1.64 & 0.53 &  7.68 & 1.15 & $0.22\mid 0.91$\\
\\
231620.78-871626.8 & 46.99121 & 10.76 & 1.408 &  1.676 & 0.009 & 0.693 & 0.569 & 1.39 & 0.81 & 4300 & 2.4 & Giant & 2.86 & 0.36 &  6.68 & 0.94 & $0.02\mid 0.82$\\
\\
\enddata
\label{table1}
\end{deluxetable}

\end{document}